\documentclass[11pt,a4paper]{article}
\pdfoutput=1

\usepackage[latin1]{inputenc}
\usepackage{amsmath}
\usepackage{amsthm}
\usepackage{amsfonts}	
\usepackage{amssymb}
\usepackage[pdftex]{color, graphicx}
\usepackage{rotating}
\usepackage{setspace}
\usepackage[mathlines]{lineno}
\usepackage{multirow}
\usepackage{pdfpages}

\usepackage[square,numbers,sort&compress]{natbib}
\usepackage[left=3cm,right=3cm,top=4cm,bottom=4cm]{geometry}

{\theoremstyle{remark}
      \newtheorem{assumption}{Assumption}}
      
{\theoremstyle{remark}
			\newtheorem{proposition}{Proposition}}

\newtheorem{theorem}{Theorem}

\title{Finite-population evolution with rare mutations in asymmetric games\footnote{Published in the \textit{Journal of Economic Theory}; doi: 10.1016/j.jet.2015.12.005.} \footnote{We are grateful to Kirill Borusyak, Drew Fudenberg, Christian Hilbe, Martin Nowak, and the referees for helpful comments. Correspondence to: carlveller@fas.harvard.edu, lhayward@math.columbia.edu.}}

\author{Carl Veller\thanks{Department of Organismic and Evolutionary Biology, Harvard University, Cambridge, MA 02138, USA} \thanks{Program for Evolutionary Dynamics, Harvard University, Cambridge, MA 02138, USA} \and Laura K.~Hayward\thanks{Department of Mathematics, Columbia University, New York, NY 10027, USA}}

\begin{document}


\maketitle

\begin{abstract}
We model evolution according to an asymmetric game as occurring in multiple finite populations, one for each role in the game, and study the effect of subjecting individuals to stochastic strategy mutations. We show that, when these mutations occur sufficiently infrequently, the dynamics over all population states simplify to an ergodic Markov chain over just the pure population states (where each population is monomorphic). This makes calculation of the stationary distribution computationally feasible. The transition probabilities of this embedded Markov chain involve fixation probabilities of mutants in single populations. The asymmetry of the underlying game leads to fixation probabilities that are derived from frequency-independent selection, in contrast to the analogous single-population symmetric-game case \cite{fudenberg2006}. This frequency independence is useful in that it allows us to employ results from the population genetics literature to calculate the stationary distribution of the evolutionary process, giving sharper, and sometimes even analytic, results. We demonstrate the utility of this approach by applying it to a battle-of-the-sexes game, a Crawford-Sobel signalling game, and the beer-quiche game of Cho and Kreps \cite{cho1987}.\\
\end{abstract}

\noindent \textit{JEL classification:} C62; C72; C73\\

\noindent \textit{Keywords:} Asymmetric games; Evolutionary dynamics; Imitation learning; Ergodic distribution

\newpage

\section{Introduction}

In evolutionary game theory, games are played within populations, and the prevalence of different strategies changes over time according to natural-selection-like dynamics \cite{maynardsmith1982, weibull1997, hofbauer1998, samuelson1998, nowak2006, sandholm2010}. This provides a natural method by which to model biological evolution \cite{maynardsmith1982} and various learning processes \cite{fudenberg1998}, and offers a `rationality-light' approach to equilibrium selection \cite{samuelson1998}. 

In the classical approach, populations are infinitely large and dynamics are deterministic; the focus is typically on the equilibrium refinement of evolutionary stability \cite{maynardsmith1982, hofbauer1998}. More recently, stochastic finite-population dynamics have been introduced into evolutionary game theory \cite{foster1990, kandori1993, young1993, nowak2006, fudenberg2006, mcavoy2015c}. These often take the form of an ergodic Markov chain---for example, when there is a positive mutation rate---the state space of which is all possible strategy compositions of the population \cite{fudenberg2006}. Ranking the various population states' weights in the stationary distribution is then a natural method of equilibrium selection \cite{foster1990, kandori1993}, and solves many problems of the deterministic approach.

A drawback is that the state space is often very large, making calculation of the stationary distribution infeasible. Addressing this, Fudenberg and Imhof \cite{fudenberg2006} study the case of a symmetric game played within a single, finite population, and show that, when the mutation rate is very small, the evolutionary process simplifies significantly. The intuition is straightforward: Starting from a pure (monomorphic) population state, we wait a very long time for a new strategy to appear in the population, because the mutation rate is small. When it does, it either goes extinct or takes over the population (`fixes'). Because this resolution of the mutant's fate occurs on a much shorter timescale than the waiting time for another mutation to occur, it typically re-establishes a pure state. The process therefore approximates a simpler process over just the pure states. This dramatic  reduction of the state space makes calculation of the stationary distribution computationally simple. 

The transition probabilities of this simpler process depend critically on the various mutants' fixation probabilities---the probability that a given strategy, having arisen in a population otherwise pure for a different strategy, subsequently fixes in that population. Because the game is symmetric, the payoffs that determine these fixation probabilities are frequency dependent---the payoff to a mutant strategy changes as its frequency in the population increases. For most evolutionary processes, frequency-dependent fixation probabilities either do not exist in closed form, or are intractable when they do \cite{nowak2006}. This significantly limits the analytical use of Fudenberg and Imhof's result. 

Here, we employ the basic machinery of Fudenberg and Imhof \cite{fudenberg2006} to derive a result similar to theirs for asymmetric games. There are several reasons why such a result is desirable. First, many situations in which we might want to study evolutionary or learning dynamics are best modelled as asymmetric games---for example, signalling games \cite{spence1973, crawford1982, grafen1990}, games of entry and entry-deterrence \cite{salop1979, milgrom1982, maynardsmith1976}, and games of time consistency and commitment \cite{kydland1977}. Second, because only strict Nash equilibria of asymmetric games are evolutionarily stable \cite{samuelson1992}, the deterministic approach based on evolutionary stability often fails. This is especially true for multi-stage asymmetric games, which typically have no strict Nash equilibria (because alternative strategies that induce the same path of play, prescribing the same actions on that path but different actions off it, are payoff equivalent).

In our model, evolution occurs in multiple interacting populations, one for each role in the underlying asymmetric game. When the mutation rate is very small, the evolutionary process simplifies to one over just the pure states (each population is monomorphic). Transition probabilities between pure states in this embedded process again depend on the fixation probabilities of single mutants, but these turn out to be much simpler than in the symmetric game case of Fudenberg and Imhof. To see this, suppose we start in a pure state. A mutant eventually arises in one of the populations, and either goes extinct or fixes in that population before another mutant arises in any of the populations. The other populations are therefore monomorphic for the duration of the mutant's extinction or fixation. But because the game is asymmetric, payoffs to different strategies in one population depend only on the states of the other populations, and so the payoffs that determine the fixation probability of the mutant (and therefore the transition probabilities in the embedded evolutionary process) are \emph{frequency independent}. Frequency-independent selection is a standard assumption in the population genetics literature \cite{crow1970, ewens2004}, and closed-form fixation probabilities (exact and approximate) exist for many evolutionary processes of interest. Using our result, we can employ these to derive sharper, and sometimes even analytical, characterizations of long run evolutionary behaviour in many asymmetric games of interest. This allows for powerful evolutionary equilibrium selection in these games.

We illustrate the utility of our result with three examples. First, in a `battle of the sexes' game, we show that a closed-form characterization of the stationary distribution is possible. Second, we consider a discrete Crawford-Sobel signalling game \cite{crawford1982}. We show that, when multiple signalling equilibria of differing information content exist for a given misalignment of signaller and receiver interests, the most informative is evolutionarily dominant. This gives a more foundational support to Crawford and Sobel's heuristic argument in favour of the most informative signalling equilibria, which they base on Schelling's \cite{schelling1960} concept of `focal points'. Finally, we apply our methodology to the `beer-quiche' game of Cho and Kreps \cite{cho1987}, and show that, while it supports the Intuitive Criterion in its selection between the two Bayesian Nash equilibria of the game, non-equilibrium states are also evolutionarily significant, especially for small population sizes.

\section{Evolution with mutations in multiple finite populations} \label{sec:setup}

Asymmetric games are characterized by the existence of multiple `roles' (`Player 1', `Player 2', etc.). In the evolutionary approach, the simplest way to incorporate multiple roles is to model evolution as occurring in multiple interacting populations \cite{maynardsmith1982, hofbauer1988, young1993, hofbauer1996, hofbauer1998, bergstrom2003, fishman2008, ohtsuki2010}.\footnote{In Section \ref{sec:discussion}, we discuss how our results relate to the alternative modelling choice of a single population, in which each generation, each member draws a role from some distribution.}

Suppose that we have an underlying game $\Gamma$ with roles $i = 1, \ldots, I$, each role associated with a finite strategy set $S_i$, and the payoff to a player in role $i$ when play is $\langle s_1, \ldots, s_I\rangle \in \prod_{i=1}^{I} S_i$ given by $\pi_i(s_1, \ldots, s_I) \in \mathbb{R}$. 

We assume the existence of $I$ populations, one for each role, with the size of each population $i$ constant through time at $N_i \in \mathbb{N}$. The overall population state at a given time is defined as the $I$-tuple of strategy frequencies in the respective populations at that time: $p^t \in \prod_{i=1}^{I} \Delta^{|S_i|}$, where $\Delta^n$ is the unit simplex in $\mathbb{R}^n$.\footnote{Since the populations are finite, $p^t$ is in fact confined to a finite subset of this space.} We shall be interested in the evolution of this population state over time.

Evolution proceeds as a stochastic process in discrete time. Each generation, each member of each population receives the expected value of interacting, according to $\Gamma$, with a group comprising one member from each other population, randomly chosen, and with each group equally likely. (The use of expected payoffs, rather than true payoffs received from single random interactions, is for the sake of tractability.) If $p_j^{t}(s_j^k)$ denotes the proportion of members of population $j$ that are playing strategy $s_j^k\in S_j$ at time $t$, then, for example, the expected payoff to a member of population $1$ who employs strategy $s^1_1 \in S_1$ in period $t$ is 
\[\mathbb{E}\pi_1(s_1^1 | p^t) = \mathbb{E}\pi_1(s_1^1 | p_{-1}^t) =  \sum_{k_2 = 1}^{|S_2|}\ldots \sum_{k_I = 1}^{|S_I|} p^{t}_2(s_2^{k_2})\ldots p^{t}_I(s_I^{k_I}) \pi_1(s_1^1, s_2^{k_2}, \ldots, s_I^{k_I}).\]
Here, $p_{-1}^t$ denotes the population states in all populations other than population 1, and signifies that the expected payoff to a strategy in population 1 depends only on the strategy frequencies in the other populations $2,\ldots, I$, a consequence of the asymmetry of the underlying game.

These expected payoffs in each population $i$ are then translated to non-negative \emph{fitnesses} $f_i(s_i^k | p_{-i})$ according to some positive monotonic transformation (possibly different for each population).\footnote{Popular choices in the evolutionary game theory literature include linear fitness, $f_i(\mathbb{E}\pi_i) = 1 + \eta_i\mathbb{E}\pi_i$, and exponential fitness, $f_i(\mathbb{E}\pi_i) = \exp(\eta_i\mathbb{E}\pi_i)$; in each case, the parameter $\eta_i > 0$ mediates the strength of selection, i.e., the sensitivity of fitness to changes in expected payoff.} In the case of no mutations, the fitnesses within each population can be used to update that population to its next-period state according to an evolutionary or imitation dynamic, usually following the general Darwinian, or `monotonicity', principle that strategies with high fitness increase in proportion relative to those with low fitness. 

Some notation: let $\mathcal{P}_i$ denote the (finite) set of all possible population states for population $i$, let $\mathcal{P} = \prod_{i=1}^I \mathcal{P}_i$ denote the set of all possible overall population states, and let $\mathcal{P}_{-i}$ denote the set of all possible population states for populations other than $i$. The set of `pure' states for population $i$, $\mathcal{P}^{\text{pure}}_i$, comprises all states in $\mathcal{P}_i$ where every member of population $i$ is playing the same strategy (in which case we say that population $i$ is `monomorphic'). Abusing notation a little, we label such states by the strategy that all members are playing, i.e., $\mathcal{P}^{\text{pure}}_i = S_i$. Finally, the set of overall pure states, $\mathcal{P}^{\text{pure}} = \prod_{i=1}^I \mathcal{P}^{\text{pure}}_i$, is the set of overall population states in which every population is pure. 

The evolutionary process with no mutations in each population $i$ is a stochastic process $\{X_i^0(t), t = 0, 1, \ldots\}$, with state space $\mathcal{P}_i$, and transition probabilities $T_i^0(p_i, p_i^\prime|p_{-i})$ for  $p = \langle p_i, p_{-i}\rangle \in \mathcal{P}, p_i^\prime \in \mathcal{P}_i$. The transition probabilities depend on the population state $p_{-i}$ because this determines fitnesses within population $i$. 

For each population $i$, we require two basic assumptions of this no-mutation evolutionary process defined by $T_i^0(p_i, p_i^\prime|p_{-i})$:\\

\begin{assumption}
If in some period a strategy in population $i$ is absent, then it is absent in all future periods. Formally, for all $\langle p_i, p_{-i}\rangle \in \mathcal{P}$, $p_i^\prime \in \mathcal{P}_i$, and $s_i \in S_i$, if $p_i(s_i) = 0$ and $T(p_i,p_i^\prime |p_{-i}) > 0$, then $p_i^\prime(s_i) = 0$.\\
\end{assumption}

\begin{assumption}
No matter the population state of other populations, any strategy currently played in $i$, unless it is played by all members of $i$, has positive probability of having increased representation next period. For any $\langle p_i, p_{-i}\rangle \in \mathcal{P}$, and for each $s_i \in S_i$ such that $0 < p_i(s_i) < 1$, there exists $p_i^\prime \in \mathcal{P}_i$ such that $p_i^\prime(s_i) > p_i(s_i)$ and $T_i^0(p_i, p_i^\prime | p_{-i}) > 0$.\\
\end{assumption}

Assumptions 1 and 2 are satisfied by many finite-population stochastic processes studied in evolutionary game theory and population genetics when there are no mutations, selection is finitely strong, and fitnesses are positive. These include stochastic models of imitation learning \cite{fudenberg1998}, the Moran process \cite{moran1958}, and the Wright-Fisher process \cite{fisher1930, wright1931}. Processes that are excluded include best-response dynamics and fictitious play \cite{fudenberg1998}.

Loosely, Assumption 1 ensures that the pure states for population $i$ are absorbing. In a learning context, it distinguishes imitation learning from other learning processes: strategies not employed by anyone in a population cannot be imitated \cite{binmore1997, fudenberg2006, sandholm2012}. It is also a natural assumption in a biological context: without mutations, the creation of novel genes, and therefore novel strategies, is not possible. 

Assumption 2 ensures that non-pure states in population $i$ are transient. This we take to be the essence of stochastic dynamics. It is important to note that the `positive probability' of Assumption 2 can be very small. It is not restrictive, for example, that unsuccessful strategies can spread in a population, since the probability that they do so can be appropriately small.

One context in which assumption 2 might appear, at first glance, to be too strong is that of imitation learning in multi-stage games, where some decision nodes are not reached given the strategies currently employed in the populations, so that `play' at these nodes cannot be directly observed.\footnote{We are grateful to a referee for emphasizing this point, and prompting the present discussion.} This is a problem for assumption 2 only if a particular condition holds, which we consider to constitute a somewhat `knife-edge' case: learning is by imitation based only on \emph{direct observation} of play.

If imitation can also be based, even if only to a very small degree, on communication between agents in a population, then actions at currently-unobserved nodes could be discussed and imitated, and so assumption 2 would be valid. This we take to be far more realistic. It is implicit in Young's \cite{young1993} assumption that `each time an agent plays he starts afresh and must ask around to find out what is going on', and a similar logic underlies many models of social learning (e.g., \cite{ellison1995}). Again, we should stress that \emph{any} amount of such communication validates assumption 2; the condition under which it is invalid is therefore a knife-edge case.\footnote{It might be objected that changing one's strategy at an unreached decision node would not alter one's payoff, so that imitative strategy changes of this sort would not be expected, but this objection fails to take into account the fundamental stochasticity of the process: even detrimental strategy changes are expected to occur with some positive probability.} Communication is especially relevant for situations where membership of the populations is not fixed through time, instead being affected by exits and entries (as modelled by birth-death processes, for example). In this case, if a decision node is currently unreached for a given population, then new entrants in that population must nonetheless have strategies that prescribe actions at the unreached nodes; in a pure imitation dynamics, they can only get these by `asking around'.

We make the further assumption that the evolutionary processes occur independently within each population, in the sense that, although the probability that population $i$ transitions from $p_i$ to $p_i^{\prime}$ between periods $t$ and $t+1$ depends on the period-$t$ population states of the other populations, the transitions that these other populations make between periods $t$ and $t+1$ do not influence the transition in population $i$. This is similar to the assumption that expected, rather than realized, payoffs are relevant for fitnesses, in the sense that it too is an abstraction from the true, random, matching of players in a given period. Like the expected payoffs assumption, it is made for tractability.

Under this assumption, the no-mutation processes $\{T_i^0\}_{i=1}^I$ aggregate to an overall no-mutation Markov process $T^0$ over the state space $\mathcal{P}$, where for $p = \langle p_1, \ldots, p_I\rangle, p^\prime = \langle p_1^\prime, \ldots, p_I^\prime\rangle \in \mathcal{P}$, $T^0(p,p^\prime) = \prod_{i=1}^I T_i^0(p_i, p^\prime_i|p_{-i})$.

We now incorporate mutations into this general evolutionary process. We specify for each population $i$ a mutation rate $\varepsilon\mu_i>0$, with $\mu_i$ a population-specific parameter that governs the between-population relative frequency of mutations, and $\varepsilon$ an across-population parameter governing the overall frequency of mutations. We then alter the above no-mutation evolutionary process as follows: From a population state $p^t$ in period $t$, a \emph{preliminary} (pre-mutation) population state $p_{(0)}^{t+1}$ for period $t+1$ is chosen according to the transition probabilities $T^0$, i.e., according to the no-mutation evolutionary process.

This preliminary population state is then subjected to random mutations of the following form: in each population $i$, each member has probability $\varepsilon\mu_i$ of discarding her strategy and randomly selecting another from the strategy space $S_i$, with each strategy (including the one she just discarded) equally likely.\footnote{We can easily allow for the possibility that not all mutations between strategies within a population are equally likely; this case is discussed in Section \ref{sec:discussion}.} This mutation process is carried out independently across the members of a population, and similarly across populations, resulting in the final population state for period $t+1$, $p^{t+1}$.

The evolutionary process with mutations can be summarized by the following scheme:
\[p^t \quad \rightarrow \quad \text{selection (stochastic)} \quad \rightarrow \quad \text{mutation (stochastic)} \quad \rightarrow \quad p^{t+1}.\]

Within each population $i$, this is a stochastic process governed by the transition probabilities $T_i^\varepsilon(p_i, p^\prime_i|p_{-i})$. These individual population processes aggregate to an overall Markov process over the state space $\mathcal{P}$, defined by the transition probabilities $T^\varepsilon(p, p^\prime) = \prod_{i=1}^I T_i^\varepsilon(p_i, p^\prime_i|p_{-i})$ (because the independence of the within-population processes is not compromised by the mutations process we have defined). 

Since $\mu_i>0$ for each population $i$, there is positive probability that, from any given population state, any state can be reached in one generation (it just requires the appropriate mutations). Consequently, the evolutionary process $T^\varepsilon(p, p^\prime)$ with positive mutation rates $\mu_i$ is an ergodic Markov chain. It therefore has a unique stationary distribution, which it approaches in the long run.

In principle, this stationary distribution is analytically calculatable, but in reality, for many games of interest, the state space (all possible population states) will usually be so large that this calculation is infeasible. In general, the size of the state space is $|\mathcal{P}| = \prod_{i=1}^I \binom{N_i + |S_i| -1}{|S_i| - 1}$. In the case of just two populations, each of size 20 members, and each with 4 strategies available to its members, the size of the state space is approximately $3\times 10^{6}$: calculating the stationary distribution thus involves solving a system of about $3\times 10^{6}$ linear equations. This problem intensifies as the population sizes increase.

In the next section, we employ a theorem of Fudenberg and Imhof \cite{fudenberg2006} to show that, when the mutation rate is very small for each population ($\varepsilon \ll 1$), the stationary distribution of the evolutionary process with mutations approximates an embedded Markov process on a much-reduced state space, the set of all pure states $\mathcal{P}^{\text{pure}}$ (the size of which does not increase with increasing population size). Moreover, the asymmetry of the underlying game will render selection frequency-independent in the rare-mutations regime. This will make calculation of the transition probabilities of this embedded Markov chain much simpler than for symmetric games.

\section{The stationary distribution when mutations are rare} \label{sec:statdist}

Assumptions 1 and 2, which concern the within-population no-mutation evolutionary processes $T^0_i$, translate into the following two straightforward propositions, stated without proof, concerning the aggregate no-mutation process $T^0$:\\

\begin{proposition}
Under $T^0$, all pure population states $p\in\mathcal{P}^{\text{pure}}$ are absorbing.\\
\end{proposition}

\begin{proposition}
Under $T^0$, all population states $p \in \mathcal{P}\backslash \mathcal{P}^{\text{pure}}$ are transient.\\
\end{proposition}

Label pure population states by $s = \langle s_1, \ldots, s_I \rangle \in \mathcal{P}^{\text{pure}}$: here, all members of population $i$ play strategy $s_i \in S_i$. Denote by $s/s_i^\prime$ the population state where every population $j\neq i$ is monomorphic for the strategy $s_j$, and population $i$ is monomorphic for the strategy $s_i$ except for one individual, who plays $s_i^\prime \neq s_i$. Let the set of all such states be $\mathcal{P}^{\text{pure}/i}$.\\

\begin{proposition}
Fix $s \in \mathcal{P}^{\text{pure}}$, and consider the limit $\lim\limits_{\varepsilon \to 0} \frac{T^{\varepsilon}(s,\, p)}{\varepsilon}$ for states $p \in \mathcal{P}\backslash\{s\}$. This limit exists for all states $p \in \mathcal{P}\backslash\{s\}$. However, $\lim\limits_{\varepsilon \to 0} \frac{T^{\varepsilon}(s,\, p)}{\varepsilon} > 0$ if, and only if, $p \in \mathcal{P}^{\text{pure}/i}$ for some $i$. Otherwise, $\lim\limits_{\varepsilon \to 0} \frac{T^{\varepsilon}(s,\, p)}{\varepsilon} = 0$.\\
\end{proposition}

To prove this, note that $T^{\varepsilon}(s,\, p)$ is a polynomial in $\varepsilon$ for all $p$. For $T^{\varepsilon}(s,\, s/s_i^\prime)$, this polynomial has lowest-order term $\frac{N_i\mu_i}{|S_i|}\varepsilon$, and so $\lim\limits_{\varepsilon \to 0} \frac{T^{\varepsilon}(s,\, s/s_i^\prime)}{\varepsilon} = \frac{N_i\mu_i}{|S_i|} > 0$. On the other hand, if the states $s$ and $p$ differ by the strategy played by more than one individual, then a one-step transition from the former to the latter requires more than one mutation, so $T^{\varepsilon}(s,\, p)$ has lowest-order term of order $\varepsilon^k$, $k\geq 2$. Thus, for such states $p$,  $\lim\limits_{\varepsilon \to 0} \frac{T^{\varepsilon}(s,\, p)}{\varepsilon} = 0$.

Proposition 3 states that mutations from a pure state to a state where only one individual in one of the populations deviates from the pure state are, for small mutation rates, at least an order of magnitude more likely than other transitions from the pure state (and, owing to the pure states being absorbing under the no-mutation process, mutations are the only way to transition out of pure states).

Now suppose that, from a pure state $s$, the system transitions to the state $s/s_i^\prime$. Since interior states in population $i$ are transient, absent further mutations in population $i$, the process will absorb either back into the pure state $s$ (the mutant strategy $s_i^\prime$ has `gone extinct') or into the pure state $\langle s_i^\prime, s_{-i}\rangle = \langle s_1, \ldots, s_{i-1}, s_i^\prime, s_{i+1}, \ldots, s_I\rangle$ (the mutant strategy $s_i^\prime$ has `fixed'). 

But when the mutation rates are very small, we should expect this extinction or fixation of strategy $s_i^\prime$ to occur before another mutant appears in population $i$, and indeed before a mutant subsequently appears in any other population. This latter fact, that no mutant is expected to appear in any of the other populations during the extinction/fixation event in population $i$, is key in determining the probability that fixation of $s_i^\prime$ will occur in population $i$. Because the underlying payoffs to, and thus fitnesses of, strategies $s_i$ (the `incumbent' strategy) and $s_i^\prime$ (the `mutant' strategy) depend only on the population states in the other populations, and since these are fixed at $s_{-i}$ for the duration of the extinction or fixation of $s_i^\prime$, the fitness difference between $s_i$ and $s_i^\prime$ is constant for the duration of this event. Thus, selection is \emph{frequency-independent} in this regime, a fact that will make the calculation of the various fixation probabilities significantly simpler.

To formalize this intuition, for states $s \in \mathcal{P}^{\text{pure}}$ and $s/s_i^\prime \in \mathcal{P}^{\text{pure}/i}$ define $\hat{\mu}_i(s_i,s_i^\prime) := \lim\limits_{\varepsilon \to 0} \frac{T^{\varepsilon}(s,\, s/s_i^\prime)}{\varepsilon} = \frac{N_i\mu_i}{|S_i|} = \hat{\mu}_i$, and let $\rho_i(s_i, s_i^\prime|s_{-i})$ be the `fixation probability' that, given that populations $-i$ remain monomorphic for strategies $s_{-i}$, a $s_i^\prime$ mutant who appears in population $i$ that is otherwise monomorphic for strategy $s_i$ subsequently fixes. Assumption 2 ensures that this probability is always positive.

Now let $K = |\mathcal{P}^{\text{pure}}| = \prod_{i=1}^I |S_i|$, and let $1, \ldots, K$ be some enumeration of the pure population states.\footnote{A particular enumeration that we have found useful: writing $K_i = \prod_{j=i+1}^I |S_i|$, enumerate the pure state $\langle s_1^{m_1}, \ldots, s_I^{m_I}\rangle$ by $\sum_{i=1}^{I-1} K_i(m_i-1) + m_I$. The population states in the pure state enumerated $n$ can then be recovered as follows: $m_I -1 = n\mod |S_I|$, and $m_i -1 = \left\lfloor\frac{n}{K_i}\right\rfloor\mod |S_i|$ for each $i < I$.} Construct a $K \times K$ transition probability matrix $\Lambda$ as follows:
\begin{itemize}
\item If the pure states labelled $m$ and $n$ are $s = \langle s_i, s_{-i} \rangle$ and $\langle s_i^\prime, s_{-i}\rangle$ respectively (i.e., pure states that differ by only one population's strategy), then $\Lambda_{mn} = \hat{\mu}_i\rho_i(s_i, s_i^\prime|s_{-i})$.
\item If the pure states $m$ and $n$ differ by more than one population's strategy, $\Lambda_{mn} = 0$.
\item Having thus defined $\Lambda_{mn}$ for all distinct $m$ and $n$, define $\Lambda_{mm} = 1 - \sum_{n\neq m} \Lambda_{mn}$.
\end{itemize}
$\Lambda$ is the transition probability matrix for a homogeneous Markov chain over the (finite) state space $\mathcal{P}^{\text{pure}}$.\footnote{If there are some $m$ such that $\Lambda_{mm} < 0$ in the above construction of $\Lambda$, one can rescale all mutation rates $\mu_{i}$ by an appropriately small factor to render all $\Lambda_{mm} > 0$. Any such rescaling will result in the same stationary distribution induced by $\Lambda$.} Moreover, this Markov chain is irreducible, since any pure state can be reached from any other with positive probability in a number of steps equal to the number of populations on whose strategies the two pure states differ.

This establishes the final proposition that we require, that $\Lambda$ induces a unique stationary distribution on the state space of pure population states \cite{karlin1975}:\\
\begin{proposition}
There is a unique vector $\lambda = (\lambda_1, \ldots, \lambda_K)$ such that $\lambda_j \geq 0$ for all $j$, $\lambda_1 + \ldots + \lambda_K = 1$, and $\lambda \Lambda = \lambda$.\\
\end{proposition}

We are now in a position to state our main result. Propositions 1-4 ensure that $T^0$, $T^\varepsilon$, and $\Lambda$ satisfy Assumptions 6-9 of Fudenberg and Imhof \cite{fudenberg2006}. Employing their Theorem 2,\footnote{A simple proof of a generalization of Fudenberg and Imhof's \cite{fudenberg2006} result, holding for more general evolutionary processes, has recently been given by McAvoy \cite{mcavoy2015a}.} we arrive at the following theorem.

\begin{theorem}
For each $\varepsilon$, denote by $\lambda^\varepsilon$ the unique stationary distribution of the Markov process $T^\varepsilon$. If $n$ corresponds, in the enumeration of pure states, to the pure state $s$, then
\[\lim_{\varepsilon \to 0} \lambda^\varepsilon(s) = \lambda_n.\]
\end{theorem}
That is, the stationary distributions of $T^\varepsilon$ approach $\lambda$ as mutation rates become small.

\section{The usefulness of the result} \label{sec:examples}

Our result is useful on two fronts. First, it extends to asymmetric games our ability to compute the limiting stationary distributions of finite-population evolutionary or imitation processes. We have argued that it is these games, and asymmetric multi-stage games in particular, for which stochastic finite-population dynamics are most relevant.

Second, the fixation probabilities used to calculate $\Lambda$ derive from frequency-independent selection. Since frequency-independent selection has long been a standard assumption of the population genetics literature, we can make use of the many results about fixation probabilities in that literature. This bridge between evolutionary game theory and classical population genetics could allow for analytical calculation of the rare-mutations stationary distribution, where this would be impossible or infeasible in a single-population symmetric game setup \cite{fudenberg2006} (where the fixation probabilities that compose $\Lambda$ derive from frequency-dependent selection, and therefore typically do not exist in closed form, or are intractable when they do \cite{nowak2006}\footnote{In the particular case where the evolutionary process is a birth-death process \cite{karlin1975, nowak2006}, a closed form for fixation probabilities exists---see, e.g., \cite[Sec.~4.7]{karlin1975} and \cite[Eq.~6.13]{nowak2006}. In general, it is a complicated expression involving the relative fitnesses of the incumbent and mutant strategies at each possible intermediate frequency of the mutant strategy. One case for which this expression simplifies to a tractable form is the Moran process, if fitnesses are calculated as exponential functions of game payoffs \cite{traulsen2008, cooney2015}. For some other frequency-dependent evolutionary processes, fixation probabilities can be shown to approach tractable representations as the population size becomes very large---see, e.g., \cite{fudenberg2008}. However, as we shall discuss in Section \ref{sec:discussion}, the cases to which the `rare mutations' approximation studied here best applies are specifically those where population sizes are not too large.})

To illustrate this, we consider three examples: a `battle of the sexes' game, a discrete Crawford-Sobel signalling game \cite{crawford1982}, and the `beer-quiche' game of Cho and Kreps \cite{cho1987}.

\subsection*{Example 1: Battle of the sexes}

The well-known `battle of the sexes' game involves a man and a woman hoping to coordinate their weekend activities, which are either going to a ballet performance (the woman's preference) or going to a rugby match (the man's preference). Both the man and the woman prefer coordination on either activity to not coordinating. The simple example we shall study is summarized by the payoff matrix:
\[\begin{array}{lc}
\\
\\
\multirow{2}{*}{\text{Man}} & B\\
 & R
\end{array} \begin{array}{cc}
\multicolumn{2}{c}{\text{Woman}}\\
B&R\\
1,2 & 0,0\\
0,0 & 2,1
\end{array}\]

To cast this into an evolutionary model, assume two separate populations of men and women, of size $N_m$ and $N_w$ respectively. Each period, each member of each population goes either to the ballet or to the rugby, and receives his/her expected payoff from interacting with a random member of the other population. (This corresponds to members of each group preferring to be at an event attended by many members of the other group, though males would prefer this to be at the rugby, and females would prefer it to be at the ballet---not too unworldly a scenario!) 

Expected payoffs $\mathbb{E}\pi$ within both populations translate to fitnesses via the linear transformation $f_\theta(\mathbb{E}\pi) = 1+\eta_\theta\mathbb{E}\pi$, where $\theta$ is either $m$ (`man') or $w$ (`woman'), and $\eta_m$ and $\eta_w$ are the strengths of selection in the men's and women's populations respectively. The evolutionary or imitation dynamics within each population are assumed to be a Moran process \cite{moran1958, nowak2006}, occurring with mutations in the manner set out in Section \ref{sec:setup}. The per-person mutation, or experimentation/error, rates in the men's and women's populations are $\varepsilon\mu_m$ and $\varepsilon\mu_w$ respectively.

In populating the rare-mutations Markov matrix $\Lambda$, we need only consider transitions between pure states where either the male population's strategy is different or the female population's strategy is different, but not both. For example, consider the transition from the pure state where the men all go to the rugby and the women all to the ballet, $(\mathbf{R},\mathbf{B})$, to the pure state where everyone goes to the rugby, $(\mathbf{R},\mathbf{R})$. (The bold font indicates that these are population strategies.) In the former `incumbent' female population, members all had fitness $1+\eta_w(0) = 1$. A mutant woman going instead to the rugby has fitness $1+\eta_w(1) = 1+\eta_w$, and thus has (frequency-independent) selective advantage $\eta_w$ over the ballet-going members of the female population. 

This frequency independence allows us to make use of the well-known formula for fixation probability under a Moran process \cite{nowak2006}: If a single mutant has selective advantage $s$ over the other members of the (size $N$) population, then its fixation probability is 
\[\rho(s) = \frac{1-1/(1+s)}{1-1/(1+s)^N}\]
for $s\neq 0$, and $\rho(0) = 1/N$. The corresponding formula for the case of frequency-dependent selection is significantly more complicated \cite{karlin1975, nowak2006}. 

The entry of $\Lambda$ corresponding to the transition $(\mathbf{R},\mathbf{B}) \rightarrow (\mathbf{R},\mathbf{R})$ is therefore
\[\Lambda_{(\mathbf{R},\mathbf{B}) \rightarrow (\mathbf{R},\mathbf{R})} = \hat{\mu}_w \rho_w(\eta_w) =  \frac{N_w \mu_w}{2} \frac{1-1/(1+\eta_w)}{1-1/(1+\eta_w)^{N_w}}.\]

For the reverse transition $(\mathbf{R},\mathbf{R}) \rightarrow (\mathbf{R},\mathbf{B})$, mutant women have fitness 1, while incumbents have fitness $1+\eta_w$. Mutants are thus at relative selective disadvantage $\frac{(1) - (1+\eta_w)}{1+\eta_w} = -\eta_w/(1+\eta_w)$, and the relevant entry of $\Lambda$ is
\[\Lambda_{(\mathbf{R},\mathbf{R}) \rightarrow (\mathbf{R},\mathbf{B})} =  \hat{\mu}_w\rho_w\left(\frac{-\eta_w}{1+\eta_w}\right) = \frac{N_w \mu_w}{2}\frac{1-(1+\eta_w)}{1-(1+\eta_w)^{N_w}}.\]

The other entries of $\Lambda$ are calculated similarly. Enumerating the pure states $(\mathbf{B},\mathbf{B})$, $(\mathbf{B},\mathbf{R})$, $(\mathbf{R},\mathbf{B})$, and $(\mathbf{R},\mathbf{R})$ as 1, 2, 3, and 4 respectively, 
\[\Lambda = \left(\begin{array}{cccc}
1 - \ldots & \hat{\mu}_w\rho_w(\frac{-2\eta_w}{1+2\eta_w}) & \hat{\mu}_m\rho_m(\frac{-\eta_m}{1+\eta_m}) & 0\\
\hat{\mu}_w\rho_w(2\eta_w) & 1 - \ldots & 0 & \hat{\mu}_m\rho_m(2\eta_m)\\
\hat{\mu}_m\rho_m(\eta_m) & 0 & 1 - \ldots & \hat{\mu}_w\rho_w(\eta_w)\\
0 & \hat{\mu}_m\rho_m(\frac{-2\eta_m}{1+2\eta_m}) & \hat{\mu}_w\rho_w(\frac{-\eta_w}{1+\eta_w}) & 1 - \ldots
\end{array}\right),\]
where the ellipses abbreviate that the rows must each sum to one.

We have a number of free parameters in this model; to wit: the sizes of, selection strengths in, and mutation rates in the two populations. As an example, suppose we set the sizes of, and selection strengths in, the two populations equal at $N$ and $\eta$ respectively. Making use of the fact that, for the Moran process, $\rho(s)/\rho(-s/[1+s]) = (1+s)^{N - 1}$, we calculate the stationary distribution of the Markov chain defined by $\Lambda$:
\[\lambda = \left[1, \frac{1}{(1+2\eta)^{N-1}}, \frac{1}{(1+\eta)^{N-1}}, 1\right]/\bar{\lambda},\]
where $\bar{\lambda}$ is a normalization constant. Notice that, in the rare mutations limit, the mutation rates, though possibly different in the two populations, do not affect the long-term distribution of states.

The proportions of time the populations spend both at the rugby and both at the ballet are equal, and are higher for larger values of the common selection strength $\eta$ and population size $N$. The intuition for this effect of $\eta$ is straightforward: A higher $\eta$ increases the fixation probabilities of positively selected mutants, and decreases the fixation probabilities of negatively selected mutants. In this coordination game, the former are always mutants leading towards the coordination equilibria $(\mathbf{B},\mathbf{B})$ and $(\mathbf{R},\mathbf{R})$, while the latter are always mutants leading away from these coordination equilibria. 

The effect of population size can most easily be seen from the ratio of transition probabilities from a non-coordination to a coordination state (positive selection $s$) and vice-versa (negative selection $-s/[1+s]$): $\rho(s)/\rho(-s/[1+s]) = (1+s)^{N - 1}$. This ratio increases with $N$, and so, for each path into and out of a coordination equilibrium, a higher $N$ increases the transition probability into, relative to the symmetric probability out.

\subsection*{Example 2: Crawford-Sobel signalling}

The next game to which we apply our result is a discrete variant of a signalling game from Crawford and Sobel \cite[Sec.~4]{crawford1982}. Suppose that there are three possible states of the world, $\theta \in \{0, 1, 2\}$, with each equally likely. A signaller observes the state of the world, and sends a costless signal $s \in \{a, b, c\}$ to a receiver, who observes only the signal, and not the state of the world. Having observed the signal, the receiver makes a decision $r$. Payoffs to signaller and receiver are as follows:
\begin{linenomath}
\begin{eqnarray*}
\pi_S(\theta, r) &=& -(r-\theta-\gamma)^2,\\
\pi_R(\theta, r) &=& -(r-\theta)^2,
\end{eqnarray*}
\end{linenomath}
where $\gamma\geq 0$ is a parameter that characterizes the signaller and receiver's misalignment of interests (for every $\theta$, the receiver's optimal decision is $\gamma$ lower than the signaller would most want it to be). For simplicity, we restrict the receiver's possible decisions $r$ to the set $\{0, 0.5, 1, 1.5, 2\}$, which covers all possible optimal decisions the receiver could make given some posterior over the state space, having observed a signal. 

For all $\gamma \geq 0$, Nash equilibria exist where the signaller sends the same signal no matter the state of the world, and the receiver, observing that signal, makes decision $r = 1$. We call these equilibria `uninformative', and label them `$xxx$', since the same signal $x \in \{a,b,c\}$ is sent for each state of the world $\{0,1,2\}$. Also, for all $\gamma$, Nash equilibria exist where the signaller sends the same signal for  states $\theta = 0$ and $\theta = 2$, and a different signal for state $\theta = 1$: to all sent signals, the receiver responds with decision $1$. Since no practical (decision-changing) information is transmitted by the signaller, we also call these equilibria, labelled `$xyx$', `uninformative'.

For sufficiently low $\gamma$, there also exist `partially informative' equilibria where, for two adjacent states of the world (i.e., \{0,1\} or \{1,2\}), the signaller sends the same signal, but for the other state of the world, sends a different signal. For such `$xxy$' and `$xyy$' equilibria, these threshold values for $\gamma$ are $0.25$ and $0.75$ respectively. 

Finally, for $\gamma \leq 0.5$, there exist `informative' equilibria, where the signaller sends a different signal for each state (`$xyz$'), and the receiver makes a decision equal to the state that the signal is sent from.

A full characterization of the Nash equilibria of this game, including the receiver's responses to unsent signals required to sustain each equilibrium, is included in an appendix.

Crawford and Sobel \cite{crawford1982} argue, somewhat informally, that for a given value of $\gamma$, the most reasonable equilibria are the most informative ones possible for that $\gamma$. This, they claim, is because these equilibria are Pareto-superior to less informative equilibria, and are salient---or `focal' in Schelling's \cite{schelling1960} language---in that they are the \emph{most} informative equilibria (the other salient equilibria are the \emph{least} informative ones, but these are ruled out on the former grounds of being Pareto-inferior to the most informative equilibria). 

The methodology developed in the present paper allows us to test this equilibrium selection prediction more formally, in the context of learning by agents. Notice that none of the equilibria that are not perfectly informative is strict, so that a deterministic infinite-population approach would be of little use here, particularly for higher values of the misalignment parameter $\gamma$ (for which the informative equilibria do not exist). Instead, our finite-population approach is better-suited to this game.

We assume two populations, one of signallers and one of receivers. The size of each population is $N$. Each signaller is equipped with a response to each possible state of the world, and each receiver with a response to each possible signal. States of the world are drawn independently for each individual interaction (i.e., there is no aggregate state of the world), and fitnesses are calculated according to expected payoffs. 

Evolution within each population is assumed to be a Wright-Fisher process \cite{fisher1930, wright1931}, which has been used as a model for both biological evolution \cite{hartl2007} as well as imitation learning \cite{traulsen2009, tarnita2009}. Expected payoffs translate to fitnesses exponentially, $f(\mathbb{E}\pi) = \exp(\eta\mathbb{E}\pi)$, with selection strength $\eta$ and per-person mutation rate $\varepsilon\mu$ in both populations.

In constructing $\Lambda$, frequency-independent selection allows us to make use of the well-known `diffusion approximation' formula for the fixation probability, under the Wright-Fisher process, of a single mutant at selective advantage $s$ in a population of size $N$ \cite{kimura1962}:
\[\rho(s) \approx \frac{1-\exp{(-s)}}{1-\exp{(-Ns)}}\]
for $s\neq 0$, and $\rho(0) = 1/N$.\footnote{The diffusion approximation formula cited above is known to be very accurate \cite{ewens1963, ewens1964}, and we would expect our results to alter very little were we to use near-exact numerical approximations of the true fixation probabilities. Such numerical estimation of fixation probabilities is computationally very expensive, which further highlights the value of our result: when selection is frequency dependent, as in the single-population symmetric-game case, fixation probabilities will usually \emph{have to be} estimated numerically, whereas in our asymmetric-game case, where selection is frequency independent, we may make use of well-known exact or approximate closed-form fixation probabilities.} Again, the case of frequency-dependent selection is significantly more complicated \cite{lessard2005, imhof2006, altrock2010}.

\begin{figure}[tb]
	\centering
		\includegraphics[width=0.70\textwidth]{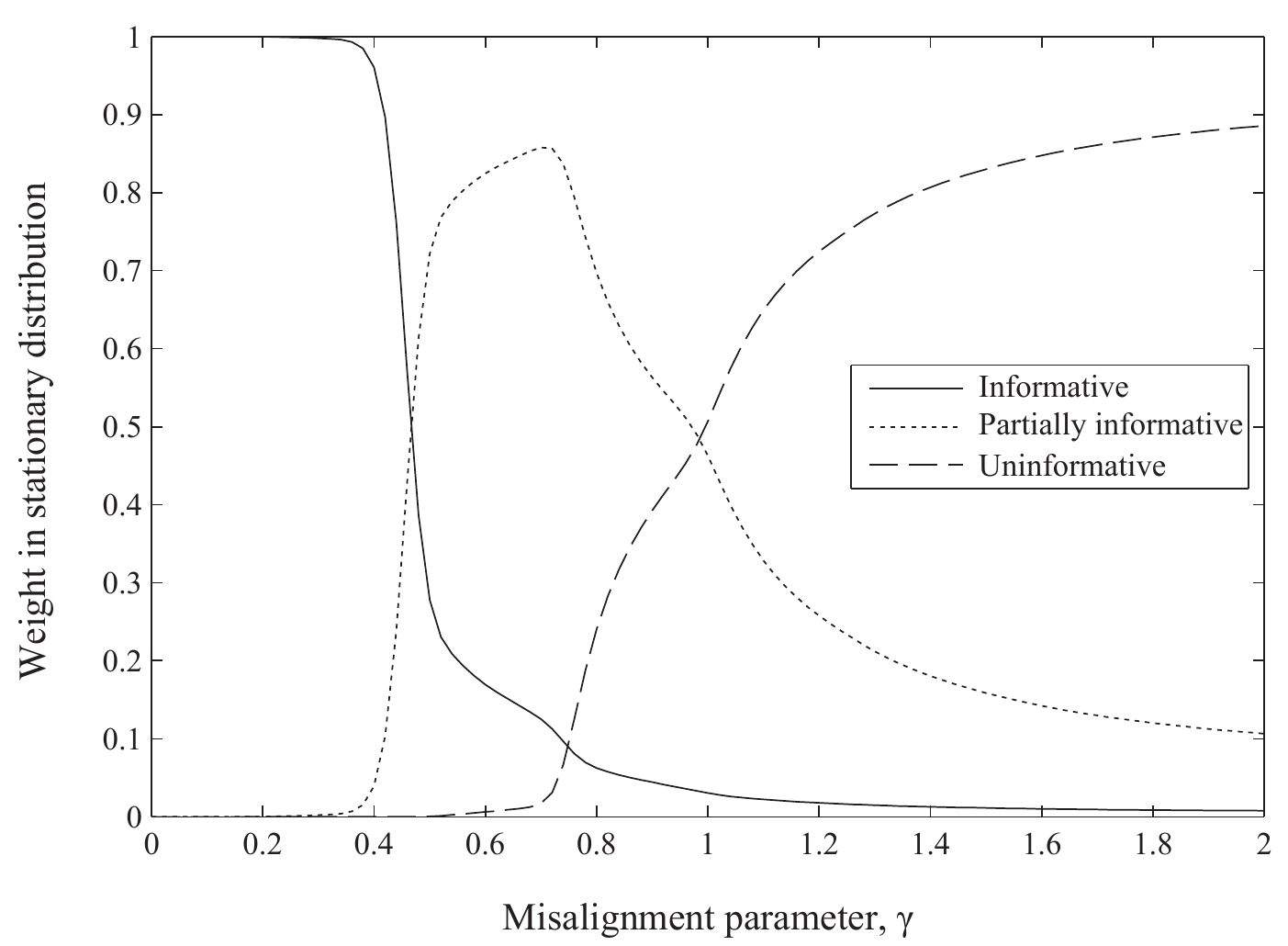}
	\caption{Frequencies of the signalling profiles of different levels of information transmission in the long-run dynamics of the Crawford-Sobel game, plotted for various values of the misalignment parameter $\gamma$. Both signaller and receiver populations are of size $N=100$; fitness is exponential in expected payoffs, with equal selection strength $\eta = 1$; mutation rates are equal in the two populations. The results are broadly consistent with Crawford and Sobel's prediction that the most informative equilibria supportable by a given value of $\gamma$ are the most reasonable for that $\gamma$.}
	\label{fig:crawford_results}
\end{figure}

We use these fixation probabilities to populate $\Lambda$ according to the method set out in Section \ref{sec:statdist}, and calculate its stationary distribution. Fig.~\ref{fig:crawford_results} plots, for the case $N = 100$ and $\eta = 1$, and for various values of the misalignment parameter $\gamma$, the relative frequencies of equilibria of different information levels in this stationary distribution.\footnote{The frequencies that we plot for a given equilibrium type are in fact those of all population states whose signalling profile is consistent with that equilibrium type: because the populations are large and selection is strong, the plotted frequencies correspond closely with those of the equilibria (which would take into account receiver behaviour too).}

It can be seen from Fig.~\ref{fig:crawford_results} that the results of the learning/evolutionary dynamics in this game broadly support Crawford and Sobel's prediction that the most informative equilibria supportable by a given $\gamma$ are the most reasonable for that $\gamma$. For low levels of misalignment $\gamma < 0.4$, the informative equilibria dominate, and information transmission is almost always perfect in the long run. For intermediate levels of misalignment ($0.4 < \gamma < 1$), partially informative equilibria, especially those of the form $xyy$, are dominant. For high levels of misalignment ($\gamma > 1$), only uninformative equilibria can be supported, and indeed such equilibria dominate the long-run dynamics.

Note that the equilibria involving signalling of the forms $xxy$ and $xyx$ do not have analogs in the equilibria of the game with continuous state, signal, and decision spaces \cite{crawford1982}; they are artefacts of the discrete structure of the game we have set up. It is reassuring, then, that they play little role in the long-run dynamics for all values of $\gamma$.

\subsection*{Example 3: The beer-quiche game}

\begin{figure}
	\centering
		\includegraphics[width=0.80\textwidth]{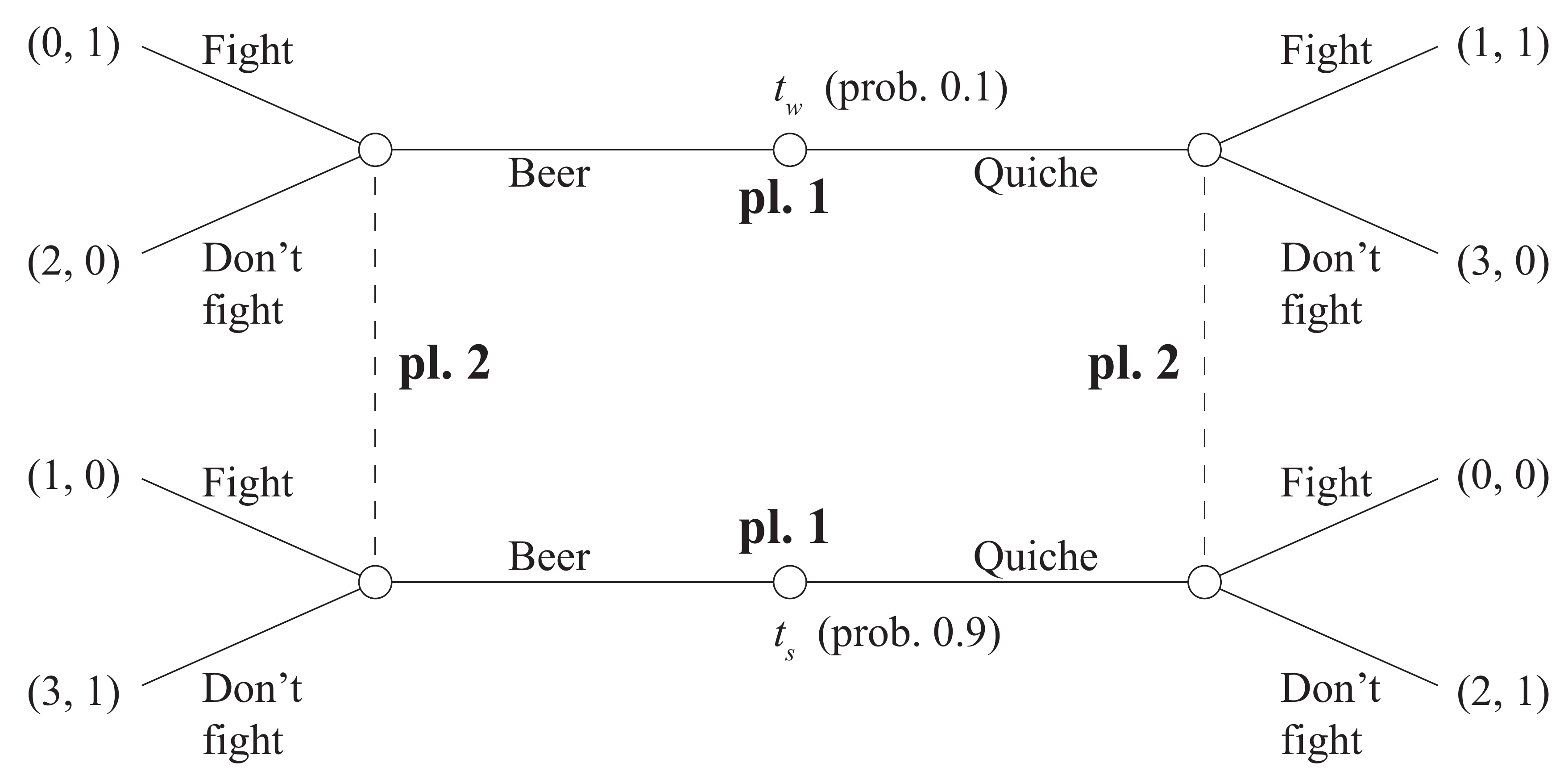}
	\caption{Extensive form setup of the beer-quiche game of Cho and Kreps \cite{cho1987}.}
	\label{fig:beerquiche_game}
\end{figure}

Our final example is the beer-quiche game of Cho and Kreps \cite{cho1987}, employed by them to illustrate the equilibrium refinement method they advance, the `Intuitive Criterion'. The extensive form of the game is given in Fig.~\ref{fig:beerquiche_game}. Player 1 is either a wimp (type $t_w$) or surly (type $t_s$), with probabilities $0.1$ and $0.9$ respectively. Player 1 knows his type; player 2 does not. Player 1 either has beer or quiche for breakfast, observed by player 2, who then chooses whether to fight player 1 or not. The payoffs are such that player 2 should choose to fight player 1 if the posterior probability he holds that player 1 is a wimp is greater than $0.5$. For any action by player 2, player 1 prefers beer for breakfast if he is surly, but quiche if he is a wimp. Regardless of player 1's type, he would prefer to avoid fighting. 

The game has two Bayesian Nash equilibria, both of the `pooling' kind: one in which player 1 eats quiche no matter his type, and one in which player 1 drinks beer no matter his type. In both cases, player 2 chooses not to fight in response to the observed behaviour of player 1, but would fight in response to the unobserved behaviour. Both pooling equilibria are sustained by player 2's `out-of-equilibrium' belief that, if he were to observe player 1 having the opposite breakfast to that consumed in equilibrium, there would be a greater-than-half chance that player 1's type was wimp. Cho and Kreps's Intuitive Criterion, however, rules out the always-quiche equilibrium, by the argument that the out-of-equilibrium beliefs that player 2 is required to hold do not survive forward-inductive reasoning \cite{cho1987} \cite[Ch.~11.2]{fudenberg1991}.

The Intuitive Criterion has been criticized as being, in some cases, too rationality-heavy \cite{fudenberg1991}. Our methodology allows us to test whether its prediction in the beer-quiche game holds up under a rationality-light learning process, where players need not even know the other players' payoffs.

We assume two populations, one for each role. Evolution proceeds in each population as a Wright-Fisher process with mutations. The population of player $i$'s, `population $i$', is of size $N_i$, with selection strength $\eta_i$, exponential fitness $f_i = \exp(\eta_i \mathbb{E}\pi)$, and per-individual mutation rate $\mu_i$. Each member of population 1 has a strategy prescribing his breakfast choice (beer or quiche) if he turns out to be wimpish (with probability $0.1$) and if he turns out to be surly (with probability $0.9$). Each member of population 2 has a strategy prescribing his response (fight or don't fight) to seeing a member of population 1 drink beer for breakfast, and to seeing a member of population 1 eat quiche. Each round, each member of each population receives his expected ex-ante (i.e., before types are chosen in population 1) payoff from interacting with a random member of the other population. 

We label pure population states by the tuple $b(t_w), b(t_s); r(B), r(Q)$: respectively, breakfast had when wimpish, breakfast had when surly; response to beer-drinking, response to quiche-eating. For the former two, $\mathbf{B}$ and $\mathbf{Q}$ represent `beer' and `quiche', while, for the latter two, $\mathbf{F}$ and $\mathbf{N}$ represent `fight' and `no fight'. Again, the bold font is used to indicate that these are population strategies.

\begin{figure}[tb]
	\centering		\includegraphics[width=0.70\textwidth]{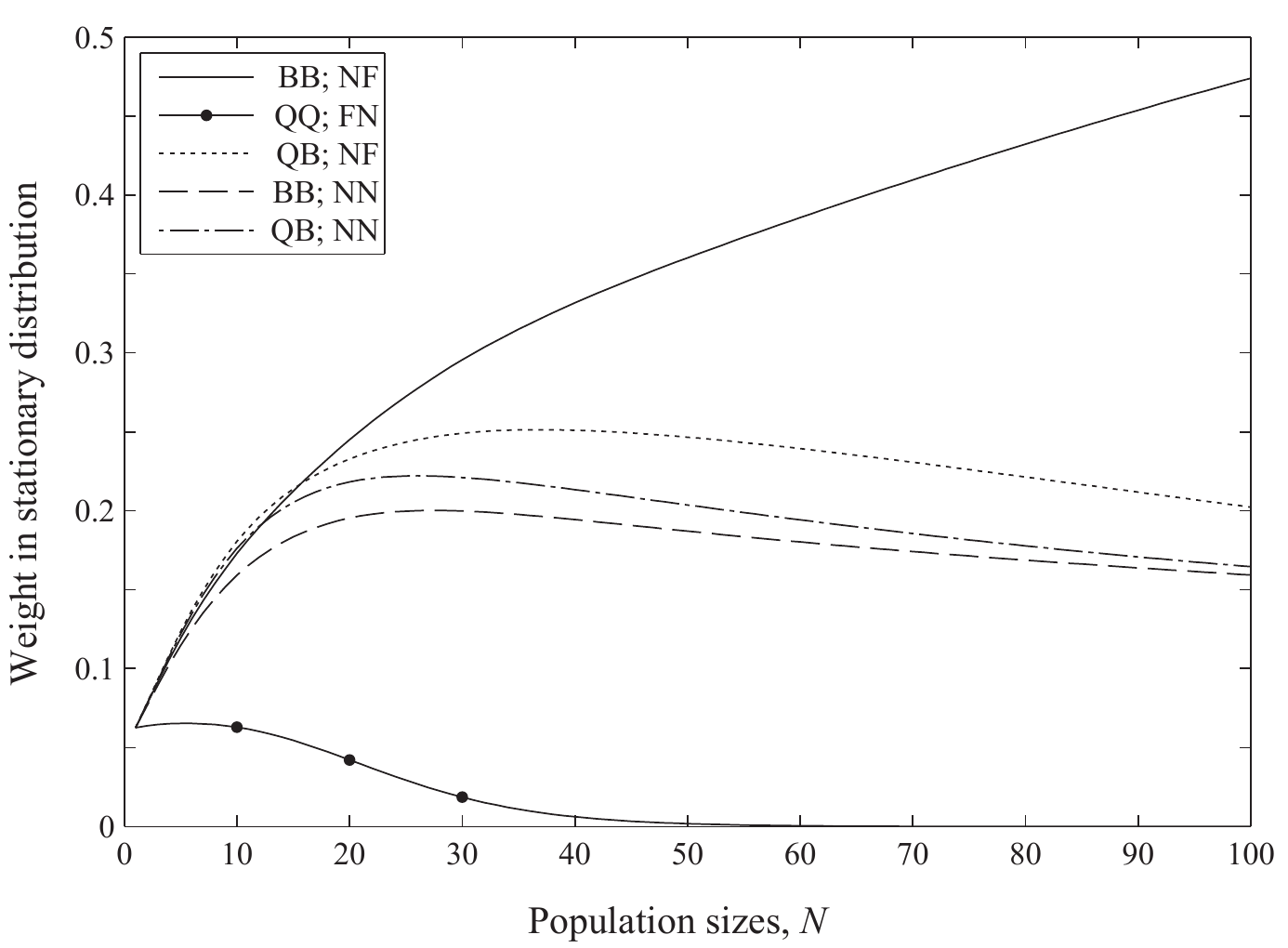}
	\caption{The frequencies of various population states in the long run dynamics of the beer-quiche game, plotted for various common population sizes $N = N_1 = N_2$. For reference, population state `$\mathbf{QB};\mathbf{NF}$' is that where members of population 1 eat quiche ($\mathbf{Q}$) if wimpish and drink beer ($\mathbf{B}$) if surly, while members of population 2 do not fight ($\mathbf{N}$) if they see beer-drinking and do fight ($\mathbf{F}$) if they see quiche-eating. The equilibrium predicted by the Intuitive Criterion, $\mathbf{BB};\mathbf{NF}$, is modal for large ($>20$), but not for low ($<20$), population sizes. The equilibrium ruled out by the Intuitive Criterion, $\mathbf{QQ};\mathbf{FN}$, is infrequent in the long-run dynamics for all population sizes.}
	\label{fig:beerquiche_results}
\end{figure}

The weights of the most popular states in the stationary distribution are displayed in Fig.~\ref{fig:beerquiche_results}, for the parameter settings $\eta_1 = \eta_2 = 0.2$, $\mu_1 = \mu_2$, and for various population sizes $N = N_1 = N_2$. For large population sizes ($N > 20$), the pooling equilibrium predicted by the Intuitive Criterion, $\mathbf{BB};\mathbf{NF}$, is the modal state in the stationary distribution. For all population sizes, the other pooling equilibrium, `all-quiche', has low weight in the stationary distribution; this supports its rejection by the Intuitive Criterion.

Apart from the fact that, of the two Bayesian Nash equilibria, the one predicted by the Intuitive Criterion is dominant, it is also of interest that non-equilibrium population states occur so frequently in the long run. These states are, in order of their weights in the stationary distribution, $\mathbf{QB};\mathbf{NF}$, $\mathbf{QB};\mathbf{NN}$, and $\mathbf{BB};\mathbf{NN}$. Indeed, for small population sizes ($N < 20$), the state $\mathbf{QB};\mathbf{NF}$ has highest weight in the stationary distribution.

The success of these non-equilibrium states is a result of neutral and nearly-neutral drift. Starting from the equilibrium state $\mathbf{BB};\mathbf{NF}$, members of population 2 who instead play $NN$ achieve the same expected payoff ($0.9$) as those playing $NF$, and so can neutrally invade the population. If they fix, the pure population state $\mathbf{BB};\mathbf{NN}$ is established. From this state, members of population 1 who play $QB$ are slightly favoured over the incumbents playing $BB$ (expected payoff $3$ versus $2.9$), and so can invade and fix, establishing the pure state $\mathbf{QB};\mathbf{NN}$. From this state, members of population 2 who play $NF$ are slightly favoured (expected payoff $1$ versus incumbent expected payoff $0.9$). If they invade and fix, pure state $\mathbf{QB};\mathbf{NF}$ is established. But from this pure state, members of population 1 who play $BB$ are slightly favoured (expected payoff $2.9$ versus incumbent $2.8$). If they invade and fix, the equilibrium pure state $\mathbf{BB};\mathbf{NF}$ is re-established. Notice that, because the reverse directions involve only neutral and slightly disfavourable mutations, they also occur with non-negligible probability, and are therefore likely to influence the stationary distribution.

The intuition for the fact that increased population size here results in the system spending more time in the Nash equilibrium state is similar to that for the same observation in the battle of the sexes. When the population size is small, mutants that are weakly selected against still have non-negligible probability of fixing, and so transitions out of $\mathbf{BB};\mathbf{NF}$ to, say, $\mathbf{QB};\mathbf{NF}$ (mutant's expected payoff only $0.1$ less than incumbents') play a role in the long-run dynamics. When the population size is very large, mutants that are weakly selected against have very little chance of fixing, and so these paths out of equilibrium are shut down, leaving only neutral paths such as $\mathbf{BB};\mathbf{NF} \rightarrow \mathbf{BB};\mathbf{NN}$. Increasing selection strength $\eta$ would have the same effect.

\section{Discussion} \label{sec:discussion}

Our model involves a number of assumptions and simplifications, four major ones of which we discuss below: (i) the assumption of multiple populations, (ii) that, in the no-mutations process, only pure states are absorbing, (iii) that mutations can, in reality, be sufficiently rare for the evolutionary dynamics to behave like the limiting case, and (iv) that mutation rates within populations are uniform. Thereafter, we briefly discuss the relevance of the approach developed in this paper for mixed-strategy equilibria.

On (i), an alternative approach would be to model evolution as occurring in a single population, wherein each agent has a strategy for every role \cite{mcavoy2015b}. Expected payoffs to players could then be computed on the basis of random assignment of roles each period.

In most learning contexts, the multiple-population setup seems more natural: we think of roles as being assigned at the outset, with each agent subsequently learning how best to play her assigned role. An example is the battle of the sexes game studied in Section \ref{sec:examples}, where the gender of each agent is fixed for the duration of his/her learning period. The multiple-population setup is also better suited to modelling the genetical evolution of multiple interacting, though reproductively distinct, species. In the context of genetical evolution within a single species, however, the more natural model is a single population in whose genomes strategies for different roles are encoded at different loci. Strategies are then collections of alleles, one for each locus, and are inherited intact (ignoring genetic recombination). In the course of the propagation of a strategy, which locus is relevant will change from generation to generation, as different roles are taken on (carrier is male or female, carrier is the incumbent occupant of a territory or the trespasser, etc.). 

When should we expect the evolutionary dynamics under this single-population model to resemble those under our multiple-population model (where each locus, or role, is treated as a separate `population')? Here, the answer is simpler for deterministic infinite-population dynamics. If there is variation within the population for alleles (/strategies) at multiple loci (multiple loci exhibit `polymorphism'), then the multiple-population approach and the single-population approach yield equivalent dynamics under the deterministic replicator dynamics if a simple condition concerning allele frequencies holds \cite{cressman2003}. This condition, known in the population genetics literature as \emph{linkage equilibrium}, amounts to statistical independence of allele frequencies across loci, and is preserved through time under the replicator dynamics \cite{cressman2003}.

In a finite population, polymorphism at multiple loci will be common if the mutation rate or the population size are sufficiently large. The stochastic nature of the evolutionary process in a finite population ensures that linkage equilibrium will not always hold, and so a `dynamical equivalence' result such as that described above is not possible. Nonetheless, if mutations at different loci occur independently, then in the regime of rare mutations studied in this paper, it will almost always be the case that at most one locus is polymorphic in the population. Thus linkage equilibrium will almost always hold, since linkage disequilibrium between two loci requires that both loci be polymorphic. In the rare mutations limit, therefore, the dynamics are the same whether we model evolution as occurring in multiple populations of loci, or in a single multi-locus population.

On (ii), we noted that, in multi-stage games, play at unreached decision nodes could not be imitated if imitation were a learning process based only on direct observation of play. Under this condition, interpreting assumption 2 as applying to imitation processes seems unjustified---mixed population states in which one population is polymorphic for an action at an unreached decision node could be maintained in perpetuity. As argued in Section \ref{sec:setup}, this condition represents an unrealistic `knife-edge' case. In reality, we expect agents to communicate about their strategies. The case of imitation only by direct observation is nonetheless a useful benchmark from which to discuss the influence of the general factors that validate assumption 2 for imitation learning processes. Despite the obvious importance of such a discussion, it has not, to our knowledge, explicitly appeared in the literature on stochastic learning in extensive form games.

On (iii), how rare do mutations have to be for the population dynamics to resemble those in our limiting case? A simple heuristic may be derived as follows: Assume all $I$ populations to be of size $N$, with a common individual mutation rate of $\mu$. Consider the case where, starting from a monomorphic state, a mutant appears in one of the populations. Under most commonly studied population dynamics (e.g., Wright-Fisher, Moran), the time that it takes this mutant either to go extinct or fix in its population is of order $N$ or less \cite{kimura1969, ewens2004}. Say that this time is $aN$. Then the probability that another mutant appears during the extinction/fixation of this mutant is about $\mu NI\times aN$; for this probability to be below some small threshold $\nu$, we require $\mu < \nu/(aIN^2)$. If this holds, the dynamics should resemble those for the limiting case $\mu \to 0$.

The bound could probably be loosened for most games, as it can be in the single-population symmetric game case (Wu et al.~\cite{wu2012}): the analogous loosening of the bound in Wu et al.~\cite{wu2012} is from order $1/N^2$ (our heuristic) to order $1/(N\ln N)$. In their case, this holds for all games except coexistence games, in which mixed equilibria are stable. The reason for this latter fact is that, if selection is very strong in a coexistence game, the population can stabilize around the mixed equilibrium for a very long period of time, long enough for another mutation to occur with non-negligible probability. In the case of coexistence games, the bound must be tightened to order $N^{-1/2}e^{-N}$ \cite{wu2012}. In asymmetric games, an analogous `negative feedback' issue could arise in a situation where two populations stabilize each other at respective mixed equilibria. A good example is the `matching pennies' game, where the row player and column player have the same strategy space (heads and tails); if the strategies played match, the row player gets payoff $+1$ and the column player gets $-1$, and if they don't match, the row player gets $-1$ and the column player $+1$. In our multi-population context, if the column population predominantly plays heads, the row population moves to predominantly playing heads, which in turn leads to a decrease in the play of heads in the column population, and a subsequent decrease in the play of heads in the row population, etc. If selection is strong enough, this situation could persist for long enough that the chance of another mutation occurring would be non-negligible. (Recent derivations \cite{sekiguchi2015} of fixation probabilities in a Moran process when multiple populations are polymorphic should be useful in analyzing such cases.) In this case, a strengthening of the `rare mutation' bound would be required: following the analysis of Wu et al.~\cite{wu2012}, we would expect a bound of order $I^{-1}N^{-1/2}e^{-N}$ to suffice.  

In any case, it is clear that our rare-mutations result is most relevant either if mutations (or experimentation and errors) occur at a low per-period rate, or if the populations under study are small, or both. In learning dynamics, interpretation of this `rare mutations' condition is difficult, since the rate of mutations is calibrated to the timescale over which strategy revisions are made. Thus, a `generation' might in fact constitute a very short period of time, and we might expect experimentation or errors to be very infrequent on such a timescale. Interpretation of this condition is easier for genetical evolution, where the timescale is in generations, and the probabilities of mutations can be reasonably well measured. For example, the point mutation rate at a single nucleotide site in humans (though known to vary across the genome \cite{wolfe1989, williams2000, smith2002}, and between the sexes \cite{hurst1998}) is of order about $10^{-8}$ per generation \cite{roach2010, lipson2015}. If we set a threshold of $\nu = 0.05$ and $a = 1$, and consider evolution at two independent loci (`roles'), then the bound $\mu < \nu/(aIN^2)$ holds for populations of up to about 1500 individuals. 

On (iv), it may be objected that, in our model, mutation rates within populations are uniform: a mutation from any strategy to any other strategy is equally likely. While this assumption may be valid in certain genetic contexts, in a learning context we might expect certain errors, or examples of experimentation, to be less likely than others \cite{fudenberg1998}. Also, in a genetical context, if we include in our concept of mutation the possibility of structural changes (e.g., rearrangements, translocations), or if we are interested in the evolutionary dynamics of a certain functional genotype relative to all other genotypes (grouped as one class), then asymmetric mutation rates would be natural \cite{nowak1992, mccandlish2014}.

Our result can be generalized in a straightforward way to incorporate heterogeneity in mutation rates within populations. If we denote by $\varepsilon\mu_i(s_i, s_i^\prime)$ the probability that a member of preliminary period-$t$ population $i$ currently employing strategy $s_i$ will mutate to playing $s_i^\prime$ in the finalized period-$t$ population, then the evolutionary process with mutations is a Markov chain $T^\varepsilon$. It is still the case that, for $s = \langle s_1, \ldots, s_I\rangle\in\mathcal{P}^{\text{pure}}$, $\lim\limits_{\varepsilon\to 0}\frac{T^\varepsilon(s, p)}{\varepsilon} = 0$ if $p \notin \mathcal{P}^{\text{pure}} \cup \mathcal{P}^{\text{pure}/i}$ for some $i$. But now, for $s/s_i^\prime \in \mathcal{P}^{\text{pure}/i}$, $\lim\limits_{\varepsilon\to 0}\frac{T^\varepsilon(s, s/s_i^\prime)}{\varepsilon} = N_i\mu_i(s_i, s_i^\prime)$. The transition probability matrix $\Lambda$ is then constructed as before.

If it is always the case that $\mu_i(s_i, s_i^\prime) > 0$, then the Markov chain defined by $\Lambda$ is irreducible, and an analogous form of Theorem 1 goes through as before. If, however, we allow there to be some $i$, $s_i$, and $s_i^\prime$ such that $\mu_i(s_i, s_i^\prime) = 0$, then the $T^\varepsilon$ are no longer guaranteed to induce irreducible Markov chains. It is then required that $T^\varepsilon$ have a unique stationary distribution for each $\varepsilon > 0$, and that there exists a unique stochastic vector $\lambda$ such that $\lambda \Lambda = \lambda$, for the analogous Theorem 2 to go through \cite{fudenberg2006, mcavoy2015a}.

A final point concerns games with mixed-strategy equilibria. In evolutionary game theory, two kinds of `mixed strategy' states must be distinguished \cite{grafen1979, bergstrom1998}. The `population kind' is where individuals within a population each play pure strategies, but different individuals play different strategies. In our setup, when mutations are rare, the system spends almost all of the long-run time in pure states (where individuals within each population all play the same strategy); mixed strategies of the `population kind' are therefore essentially never observed. The underlying reason is that these polymorphic states are transient under the no-mutations process. For a different reason, these `population kind' mixed states are also excluded by the evolutionary stability concept of infinite-population deterministic dynamics in asymmetric games: the component strategies of an equilibrium mixed state must have equal fitness, but then any of them could be involved in a `neutral invasion' of the state \cite{selten1980}.

The second kind of mixed strategy state is the `individual kind', and involves the individuals of a population all playing the same mixed strategy. Unlike the `population kind', such states can be evolutionarily stable in infinite-population dynamics. They do, however, raise a problem for our finite-population approach. Allowing individuals to play any mixed strategy requires an infinite strategy space (the unit simplex in $\mathbb{R}^{|S_i|}$, for population $i$ with pure-strategy space $S_i$), and therefore an infinitely large state space. A workaround would be to approximate the infinite strategy space $\mathbb{R}^{|S_i|}$ by a discrete lattice. 

\bibliography{small_mutations_cite}

\begin{thebibliography}{10}

\bibitem{fudenberg2006}
Fudenberg D, Imhof LA.
\newblock Imitation processes with small mutations.
\newblock Journal of Economic Theory. 2006;131(1):251--262.

\bibitem{cho1987}
Cho IK, Kreps DM.
\newblock Signaling games and stable equilibria.
\newblock The Quarterly Journal of Economics. 1987;102(2):179--221.

\bibitem{maynardsmith1982}
Maynard~Smith J.
\newblock {Evolution and the Theory of Games}.
\newblock Cambridge University Press; 1982.

\bibitem{weibull1997}
Weibull JW.
\newblock Evolutionary Game Theory.
\newblock MIT press; 1997.

\bibitem{hofbauer1998}
Hofbauer J, Sigmund K.
\newblock Evolutionary Games and Population Dynamics.
\newblock Cambridge University Press; 1998.

\bibitem{samuelson1998}
Samuelson L.
\newblock Evolutionary games and equilibrium selection.
\newblock MIT Press; 1998.

\bibitem{nowak2006}
Nowak MA.
\newblock Evolutionary dynamics.
\newblock Harvard University Press; 2006.

\bibitem{sandholm2010}
Sandholm WH.
\newblock Population Games and Evolutionary Dynamics.
\newblock MIT press; 2010.

\bibitem{fudenberg1998}
Fudenberg D, Levine DK.
\newblock {The Theory of Learning in Games}.
\newblock MIT Press; 1998.

\bibitem{foster1990}
Foster D, Young P.
\newblock Stochastic evolutionary game dynamics.
\newblock Theoretical Population Biology. 1990;38(2):219--232.

\bibitem{kandori1993}
Kandori M, Mailath GJ, Rob R.
\newblock Learning, Mutation, and Long Run Equilibria in Games.
\newblock Econometrica. 1993;61(1):29--56.

\bibitem{young1993}
Young HP.
\newblock The evolution of conventions.
\newblock Econometrica. 1993;61(1):57--84.

\bibitem{mcavoy2015c}
McAvoy A.
\newblock {Stochastic selection processes}.
\newblock arXiv preprint arXiv:151105390. 2015;.

\bibitem{spence1973}
Spence AM.
\newblock Job market signaling.
\newblock The Quarterly Journal of Economics. 1973;87(3):355--374.

\bibitem{crawford1982}
Crawford VP, Sobel J.
\newblock Strategic information transmission.
\newblock Econometrica. 1982;50(6):1431--1451.

\bibitem{grafen1990}
Grafen A.
\newblock Biological signals as handicaps.
\newblock Journal of Theoretical Biology. 1990;144(4):517--546.

\bibitem{salop1979}
Salop SC.
\newblock Strategic Entry Deterrence.
\newblock American Economic Review. 1979;69(2):335--38.

\bibitem{milgrom1982}
Milgrom P, Roberts J.
\newblock Predation, reputation, and entry deterrence.
\newblock Journal of Economic Theory. 1982;27(2):280--312.

\bibitem{maynardsmith1976}
Maynard~Smith J, Parker GA.
\newblock The logic of asymmetric contests.
\newblock Animal behaviour. 1976;24(1):159--175.

\bibitem{kydland1977}
Kydland FE, Prescott EC.
\newblock {Rules Rather than Discretion: The Inconsistency of Optimal Plans}.
\newblock The Journal of Political Economy. 1977;85(3):473--492.

\bibitem{samuelson1992}
Samuelson L, Zhang J.
\newblock Evolutionary stability in asymmetric games.
\newblock Journal of Economic Theory. 1992;57(2):363--391.

\bibitem{crow1970}
Crow JF, Kimura M.
\newblock An Introduction to Population Genetics Theory.
\newblock Harper \& Row; 1970.

\bibitem{ewens2004}
Ewens WJ.
\newblock Mathematical Population Genetics. I. Theoretical Introduction.
\newblock Springer; 2004.

\bibitem{schelling1960}
Schelling TC.
\newblock {The Strategy of Conflict}.
\newblock Harvard University Press; 1960.

\bibitem{hofbauer1988}
Hofbauer J, Sigmund K.
\newblock The Theory of Evolution and Dynamical Systems: Mathematical Aspects
  of Selection.
\newblock Cambridge University Press; 1988.

\bibitem{hofbauer1996}
Hofbauer J.
\newblock {Evolutionary dynamics for bimatrix games: a Hamiltonian system?}
\newblock Journal of Mathematical Biology. 1996;34(5-6):675--688.

\bibitem{bergstrom2003}
Bergstrom CT, Lachmann M.
\newblock {The Red King effect: when the slowest runner wins the coevolutionary
  race}.
\newblock Proceedings of the National Academy of Sciences.
  2003;100(2):593--598.

\bibitem{fishman2008}
Fishman MA.
\newblock Asymmetric evolutionary games with non-linear pure strategy payoffs.
\newblock Games and Economic Behavior. 2008;63(1):77--90.

\bibitem{ohtsuki2010}
Ohtsuki H.
\newblock Stochastic evolutionary dynamics of bimatrix games.
\newblock Journal of Theoretical Biology. 2010;264(1):136--142.

\bibitem{moran1958}
Moran PAP; Cambridge~Univ Press.
\newblock Random processes in genetics.
\newblock Proceedings of the Cambridge Philosophical Society.
  1958;54(01):60--71.

\bibitem{fisher1930}
Fisher RA.
\newblock {The Genetical Theory of Natural Selection}.
\newblock Clarendon Press; 1930.

\bibitem{wright1931}
Wright S.
\newblock {Evolution in Mendelian populations}.
\newblock Genetics. 1931;16(2):97--159.

\bibitem{binmore1997}
Binmore K, Samuelson L.
\newblock Muddling through: Noisy equilibrium selection.
\newblock Journal of Economic Theory. 1997;74(2):235--265.

\bibitem{sandholm2012}
Sandholm WH.
\newblock Stochastic imitative game dynamics with committed agents.
\newblock Journal of Economic Theory. 2012;147(5):2056--2071.

\bibitem{ellison1995}
Ellison G, Fudenberg D.
\newblock Word-of-mouth communication and social learning.
\newblock The Quarterly Journal of Economics. 1995;110(1):93--125.

\bibitem{karlin1975}
Karlin S, Taylor HM.
\newblock {A First Course in Stochastic Processes}.
\newblock Academic Press; 1975.

\bibitem{mcavoy2015a}
McAvoy A.
\newblock {Comment on ``Imitation processes with small mutations''
  [J.~Econ.~Theory 131 (2006) 251--262]}.
\newblock Journal of Economic Theory. 2015;159:66--69.

\bibitem{traulsen2008}
Traulsen A, Shoresh N, Nowak MA.
\newblock Analytical results for individual and group selection of any
  intensity.
\newblock Bulletin of Mathematical Biology. 2008;70(5):1410--1424.

\bibitem{cooney2015}
Cooney D, Veller C.
\newblock {Assortment and the evolution of cooperation in a Moran process with
  exponential fitness}.
\newblock arXiv preprint arXiv:150905757. 2015;.

\bibitem{fudenberg2008}
Fudenberg D, Imhof LA.
\newblock Monotone imitation dynamics in large populations.
\newblock Journal of Economic Theory. 2008;140(1):229--245.

\bibitem{hartl2007}
Hartl DL, Clark AG.
\newblock Principles of Population Genetics.
\newblock 4th ed. Sinauer; 2007.

\bibitem{traulsen2009}
Traulsen A, Hauert C.
\newblock Stochastic evolutionary game dynamics.
\newblock In: Schuster HG, editor. Review of Nonlinear Dynamics and Complexity.
  vol.~2. Wiley-VCH; 2009. p. 25--61.

\bibitem{tarnita2009}
Tarnita CE, Antal T, Ohtsuki H, Nowak MA.
\newblock Evolutionary dynamics in set structured populations.
\newblock Proceedings of the National Academy of Sciences.
  2009;106(21):8601--8604.

\bibitem{kimura1962}
Kimura M.
\newblock On the probability of fixation of mutant genes in a population.
\newblock Genetics. 1962;47(6):713--719.

\bibitem{ewens1963}
Ewens WJ.
\newblock Numerical results and diffusion approximations in a genetic process.
\newblock Biometrika. 1963;50(3 and 4):241--249.

\bibitem{ewens1964}
Ewens WJ.
\newblock The pseudo-transient distribution and its uses in genetics.
\newblock Journal of Applied Probability. 1964;1(1):141--156.

\bibitem{lessard2005}
Lessard S.
\newblock {Long-term stability from fixation probabilities in finite
  populations: New perspectives for ESS theory}.
\newblock Theoretical Population Biology. 2005;68(1):19--27.

\bibitem{imhof2006}
Imhof LA, Nowak MA.
\newblock {Evolutionary game dynamics in a Wright-Fisher process}.
\newblock Journal of Mathematical Biology. 2006;52(5):667--681.

\bibitem{altrock2010}
Altrock PM, Gokhale CS, Traulsen A.
\newblock Stochastic slowdown in evolutionary processes.
\newblock Physical Review E. 2010;82(1):011925.

\bibitem{fudenberg1991}
Fudenberg D, Tirole J.
\newblock {Game Theory}.
\newblock MIT Press; 1991.

\bibitem{mcavoy2015b}
McAvoy A, Hauert C.
\newblock Asymmetric evolutionary games.
\newblock PLoS Computational Biology. 2015;11(8):e1004349.

\bibitem{cressman2003}
Cressman R.
\newblock Evolutionary dynamics and extensive form games.
\newblock MIT Press; 2003.

\bibitem{kimura1969}
Kimura M, Ohta T.
\newblock The average number of generations until fixation of a mutant gene in
  a finite population.
\newblock Genetics. 1969;61(3):763.

\bibitem{wu2012}
Wu B, Gokhale CS, Wang L, Traulsen A.
\newblock How small are small mutation rates?
\newblock Journal of Mathematical Biology. 2012;64(5):803--827.

\bibitem{sekiguchi2015}
Sekiguchi T, Ohtsuki H.
\newblock Fixation probabilities of strategies for bimatrix games in finite
  populations.
\newblock Dynamic Games and Applications. 2015;p. 1--19.

\bibitem{wolfe1989}
Wolfe KH, Sharp PM, Li WH.
\newblock Mutation rates differ among regions of the mammalian genome.
\newblock Nature. 1989;337(6204):283--285.

\bibitem{williams2000}
Williams EJB, Hurst LD.
\newblock The proteins of linked genes evolve at similar rates.
\newblock Nature. 2000;407(6806):900--903.

\bibitem{smith2002}
Smith NGC, Webster MT, Ellegren H.
\newblock Deterministic mutation rate variation in the human genome.
\newblock Genome research. 2002;12(9):1350--1356.

\bibitem{hurst1998}
Hurst LD, Ellegren H.
\newblock Sex biases in the mutation rate.
\newblock Trends in Genetics. 1998;14(11):446--452.

\bibitem{roach2010}
Roach J, Glusman G, Smit A, Huff C, Hubley R, Shannon P, et~al.
\newblock Analysis of Genetic Inheritance in a Family Quartet by Whole-Genome
  Sequencing.
\newblock Science. 2010;328(5978):636--639.

\bibitem{lipson2015}
Lipson M, Loh PR, Sankararaman S, Patterson N, Berger B, Reich D.
\newblock Calibrating the human mutation rate via ancestral recombination
  density in diploid genomes.
\newblock PLoS Genetics. 2015;11(11).

\bibitem{nowak1992}
Nowak MA.
\newblock What is a quasispecies?
\newblock Trends in Ecology \& Evolution. 1992;7(4):118--121.

\bibitem{mccandlish2014}
McCandlish DM, Stoltzfus A.
\newblock {Modeling evolution using the probability of fixation: History and
  implications}.
\newblock The Quarterly Review of Biology. 2014;89(3):225--252.

\bibitem{grafen1979}
Grafen A.
\newblock The hawk-dove game played between relatives.
\newblock Animal Behaviour. 1979;27(3):905--907.

\bibitem{bergstrom1998}
Bergstrom CT, Godfrey-Smith P.
\newblock On the evolution of behavioral heterogeneity in individuals and
  populations.
\newblock Biology and Philosophy. 1998;13(2):205--231.

\bibitem{selten1980}
Selten R.
\newblock A note on evolutionarily stable strategies in asymmetric animal
  conflicts.
\newblock Journal of Theoretical Biology. 1980;84(1):93--101.

\end{thebibliography}

\includepdf{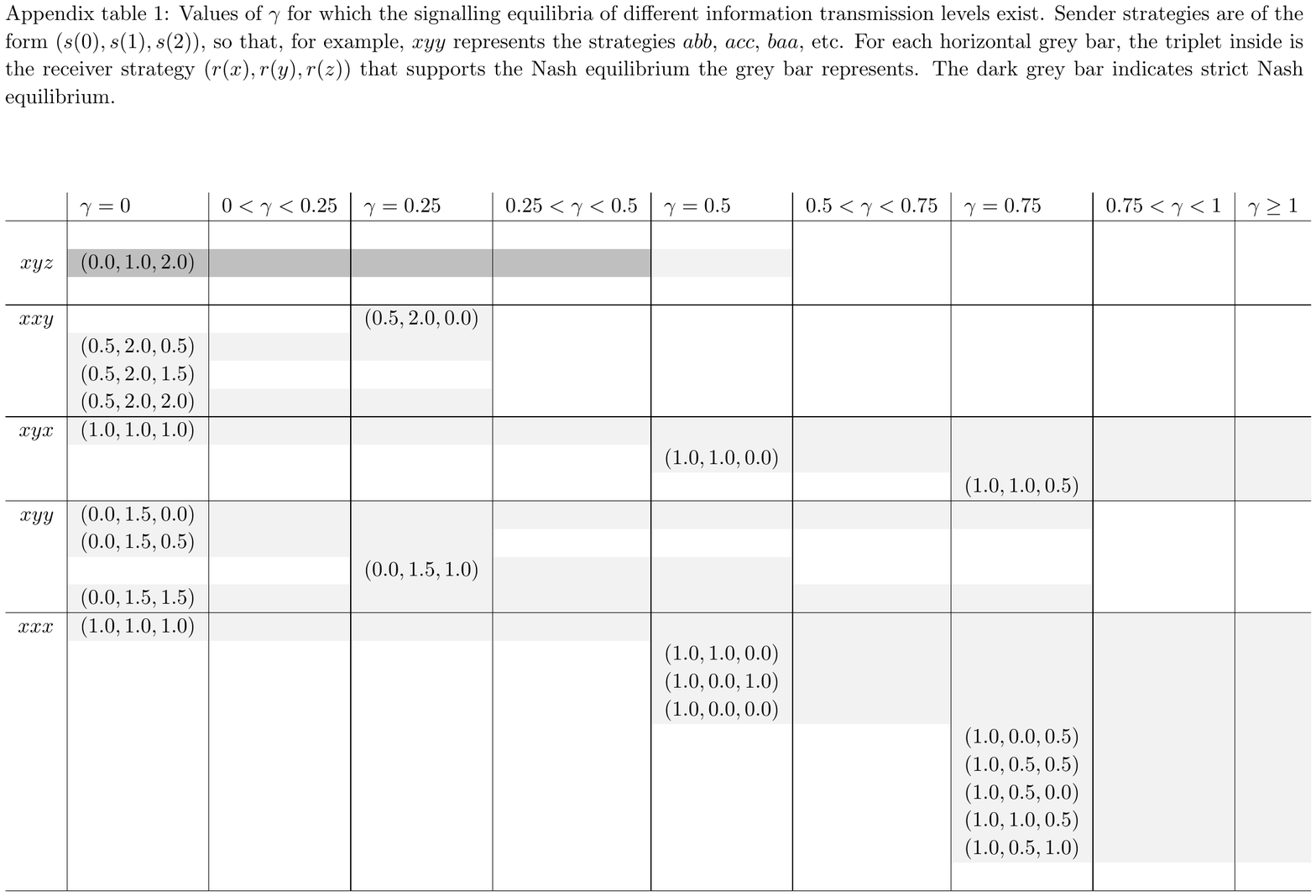}

\end{document}